\documentclass[a4paper]{jpconf}
\usepackage{graphicx}
\begin{document}
\title{Equation of state description of the dark energy transition between quintessence and phantom regimes}

\author{Hrvoje \v{S}tefan\v{c}i\'{c}\footnote{On leave of absence from the Theoretical Physics Division, Rudjer Bo\v{s}kovi\'{c} Institute, Zagreb, Croatia}}

\address{Departament d'Estructura i Constituents de la Mat\`{e}ria, Universitat de Barcelona, Av. Diagonal 647,  08028 Barcelona, Catalonia, Spain}

\ead{stefancic@ecm.ub.es}

\begin{abstract}
The dark energy crossing of the cosmological constant boundary (the transition between the quintessence and phantom regimes) is described in terms of the implicitly defined dark energy equation of state. The generalizations of the models explicitly constructed to exhibit the crossing provide the insight into the cancellation mechanism which makes the transition possible.  
\end{abstract}

The determination of the physical mechanism behind the observationally established late-time accelerated expansion of the universe \cite{data} is presently a major challenge in theoretical cosmology. Among many different approaches to the solution of this problem, the concept of dark energy plays a central role, either as a fundamental component of the universe or as an effective description of other physical mechanisms. The essential characteristics of the dark energy model are contained in the parameter of its equation of state (EOS), $p=w\rho$, where $p$ and $\rho$ denote the pressure and energy density of dark energy, respectively.

Many recent analyses of the cosmological observational data \cite{obser} obtain the transition of the dark energy from the quintessence regime ($w > -1$) to the phantom regime ($w < -1$) as a best fit description. Apart from being favoured observationally (although presently other possibilities such as the $\Lambda$CDM cosmology are consistent with the data), the cosmological constant boundary crossing is of substantial theoretical interest \cite{prethodni} since it unifies important aspects of many previously extensively studied dark energy models.  

Unlike the standard definition of the dark energy EOS in which the dark energy pressure is given as an analytical function of the energy density, $p=p(\rho)$, we define the dark energy EOS in a more general way \cite{PRDcross}. The dark energy EOS is defined parametrically, i.e. as a pair of quantities depending on the cosmic time $(p(t),\rho(t))$ or the scale factor in the expanding universe $(p(a),\rho(a))$. This definition is sufficiently general to comprise large classes of dark energy models.

We start with an explicit construction of the dark energy model which possesses the property of the CC boundary crossing. Its energy density has the following scaling  
\begin{equation}
\label{eq:denmod1}
\rho = C_{1} \left( \frac{a}{a_{0}} \right)^{-3(1+\gamma)} + 
C_{2} \left( \frac{a}{a_{0}} \right)^{-3(1+\eta)} \, ,
\end{equation}
where $\gamma > -1$ and $\eta < -1$. The parameter of its equation of state 
\begin{equation}
\label{eq:wformod1}
w=\frac{\gamma + \eta \frac{\gamma - w_{0}}{w_{0} - \eta} \left( \frac{a}{a_{0}} 
\right)^{3(\gamma - \eta)}}{1 + \frac{\gamma - w_{0}}{w_{0} - \eta} \left( \frac{a}{a_{0}} 
\right)^{3(\gamma - \eta)}} \, ,
\end{equation}
smoothly transits from $\gamma$ at $a \rightarrow 0$ to $\eta$ at $a \rightarrow \infty$. The calculation of the dark energy pressure $p$ from the dark energy energy-momentum tensor conservation and the elimination of the $(a/a_{0})^3$ term from the expressions for $\rho$ and $p$ yields the dark energy EOS
\begin{equation}
\label{eq:eosdetmod1}
\frac{p -\eta \rho}{(\gamma-\eta) C_{1}} =
\left( \frac{\gamma \rho - p}{(\gamma-\eta) C_{2}} \right)^{(1+\gamma)/(1+\eta)}
\, .
\end{equation}
The most important feature of this EOS is that it is {\em implicitly} defined. Since the model (\ref{eq:denmod1}) is explicitly constructed to yield the transition, in order to study the conditions for the occurrence of the transition, we turn to the generalized implicitly defined dark energy EOS which comprises (\ref{eq:denmod1}) as a special case, but also includes cases in which the CC boundary crossing is not possible \cite{PRDsing}. We consider
\begin{equation}
\label{eq:eosgenmod1}
A \rho + B p = (C \rho + D p)^{\alpha} \, ,
\end{equation}
where $A$, $B$, $C$, $D$ and $\alpha$ are real coefficients. This model leads to the following equation of evolution for the parameter $w$ of EOS:
\begin{equation}
\label{eq:evolw1}
\left( \frac{\alpha}{(F+w)(1+w)}-\frac{1}{(E+w)(1+w)} \right) dw = 3(\alpha - 1)
\frac{da}{a} \, ,
\end{equation}
where $E=A/B$ and $F=C/D$.
A closer inspection of this equation reveals that in the general case there exist boundaries $-E$, $-F$ and $-1$ which $w$ cannot cross at a finite value of the scale factor, but can only approach them asymptotically. However, for a special choice of parameters some of these boundaries can be removed. The solution of (\ref{eq:evolw1}) for the most interesting case $E \neq -1$ and $F \neq -1$   
\begin{equation}
\label{eq:solw}
\hspace{-1.5cm} \left| \frac{w+F}{w_{0}+F} \right|^{\alpha/(1-F)} 
\left| \frac{w+E}{w_{0}+E} \right|^{-1/(1-E)}
\left| \frac{1+w}{1+w_{0}} \right|^{1/(1-E)-\alpha/(1-F)}
= \left( \frac{a}{a_{0}} \right)^{3(\alpha-1)} \, 
\end{equation}
shows that for $\alpha=0$ the boundary at $-F$ is removed, for $\alpha \rightarrow \pm \infty$ the boundary at $-E$ does not exist, whereas for $\alpha_{{\mathrm cross}}=(1-F)/(1-E)$ the CC boundary is removed and can be crossed. The reformulation of (\ref{eq:evolw1}) in the form 
\begin{equation}
\label{eq:wstar}
\frac{w+\frac{\alpha E - F}{\alpha - 1}}{(F+w)(E+w)(1+w)} dw = 3 \frac{da}{a} \, ,
\end{equation}
provides additional insight into the mechanism of the crossing.
Namely, when $\alpha$ acquires one of the values specified above, the terms in the numerator and the denominator of the left-hand side of (\ref{eq:wstar}) get cancelled. The mechanism of the crossing of any of the boundaries in the problem (including the CC boundary) is mathematically expressed as the cancellation of the term corresponding to this boundary. 

It is also possible to construct a model in which the transition can proceed in both directions between the quintessence and phantom regime \cite{PRDcross}:
\begin{equation}
\label{eq:densitymod2}
\rho = \left( C_{1} \left( \frac{a}{a_{0}} \right)^{-3(1+\gamma)/b} + 
C_{2} \left( \frac{a}{a_{0}} \right)^{-3(1+\eta)/b} \right)^{b} \, . 
\end{equation}
%
This model is also described by the implicitly defined EOS 
\begin{equation}
\label{eq:eosdetmod2}
\frac{p -\eta \rho}{(\gamma-\eta) C_{1}} =
\rho^{((1-b)(\gamma - \eta))/(b(1+\eta))}
\left( \frac{\gamma \rho - p}{(\gamma-\eta) C_{2}} \right)^{(1+\gamma)/(1+\eta)}
\, ,
\end{equation}
whereas the corresponding generalization
\begin{equation}
\label{eq:eosmodel2}
A \rho + B p = (C \rho + D p)^{\alpha} (M \rho + N p)^{\beta} \, ,
\end{equation}
also exhibits the possibility of removing of the boundaries in the problem \cite{PRDcross}, which is controlled by the similar cancellation mechanism as for the model (\ref{eq:eosgenmod1}).
Moreover, the consideration of a nontrivial model with an EOS
\begin{equation}
\label{eq:ton}
A \rho^{2n+1} + B p^{2n+1} = (C \rho^{2n+1} + D p^{2n+1})^{\alpha} \, 
\end{equation}
reveals that this model can describe the crossing of the CC boundary for a specific choice of parameters.
Finally, let us mention that apart from the models of the CC boundary crossing in which the dark energy is a separately conserved component, there are many approaches in which the crossing is an effective phenomenon, see e.g. \cite{SiS}.

In conclusion, the implicitly defined dark energy EOS can describe the phenomenon of the CC boundary crossing. The mechanism behind the crossing is related to the cancellation of the terms corresponding to the CC boundary. In the generalized models the model parameters need to acquire special values in order to make the transition, i.e. the cancellation, possible.   

\ack

The author acknowledges the support of the Secretar\'{\i}a de Estado de Universidades e Investigaci\'{o}n of the Ministerio de Educaci\'{o}n y Ciencia of Spain within the program ``Ayudas para movilidad de Profesores de Universidad e Investigadores espa\~{n}oles y extranjeros". This work has been supported in part by MEC and FEDER under project 2004-04582-C02-01 and by the Dep. de Recerca de la Generalitat de Catalunya under contract CIRIT GC 2001SGR-00065. The author would like to thank the Departament E.C.M. of the Universitat de Barcelona for the hospitality.

\end{document}